\documentclass[useAMS,usegraphicx,usenatbib]{mn2e}

\usepackage{graphicx}
\usepackage{times} 
\usepackage{amssymb}
\usepackage{amsmath}
\usepackage{lscape}
\usepackage{url}
\newif\ifAMStwofonts
\AMStwofontstrue

\newcommand{\gtapprox}{\raisebox{-0.5ex}{$\,\stackrel{>}{\scriptstyle
\sim}\,$}}

\def\chandra{{\it Chandra}}

\def\xmm{{\it XMM-Newton}}

\def\xspec{\hbox{\it XSPEC}}
\def\xspecv{{\it XSPEC}{\rm\thinspace v\thinspace 11.3.2}}

\def\fri{\hbox{\rm FR\thinspace I}}
\def\frii{\hbox{\rm FR\thinspace II}}

\def\ks{\hbox{$\rm\thinspace ks$}}

\def\yr{\hbox{$\rm\thinspace yr$}}

\def\mhz{\hbox{$\rm\thinspace MHz$}}
\def\ghz{\hbox{$\rm\thinspace GHz$}}

\def\um{{\hbox{$\rm\thinspace \umu m$}}}

\def\kpc{\hbox{$\rm\thinspace kpc$}}
\def\mpc{\hbox{$\rm\thinspace Mpc$}}

\def\as{\hbox{$\rm\thinspace arcsec$}}

\def\pcmsq{\hbox{$\rm\thinspace cm^{-2}$}}
\def\kmpspmpc{\hbox{$\rm\thinspace km~s^{-1}~Mpc^{-1}$}}

\def\kev{\hbox{$\rm\thinspace keV$}}

\def\mjy{\hbox{$\rm\thinspace mJy$}}
\def\mjypb{\hbox{$\rm\thinspace mJy/beam$}}
\def\njy{\hbox{$\rm\thinspace nJy$}}

\def\erg{\hbox{$\rm\thinspace erg$}}
\def\ergpcmc{\hbox{$\rm\thinspace erg~cm^{-3}$}}

\def\photpkevpcmsqps{\hbox{$\rm\thinspace ct~keV^{-1}~cm^{-2}~s^{-1}$}}

\def\ergps{\hbox{$\rm\thinspace erg~s^{-1}$}}

\def\ergpcmc{\hbox{$\rm\thinspace erg~cm^{-3}$}}

\def\g{\hbox{$\rm\thinspace G$}}

\def\qc{\hbox{\rm 6C\,0905+39}}

\begin{document}

\title[] {The inverse-Compton X-ray-emitting lobes of the
  high-redshift giant radio galaxy \qc} \author[M. C. Erlund et al.]
{\parbox[]{6.in} {M.~C.  Erlund,$^{1}$\thanks{E-mail:
      mce@ast.cam.ac.uk} A.~C.
    Fabian$^{1}$ and Katherine~M. Blundell.$^{2}$} \\\\
  \footnotesize
  $^{1}$Institute of Astronomy, Madingley Road, Cambridge CB3 0HA\\
  $^{2}$University of Oxford, Astrophysics, Keble Road, Oxford OX1
  3RH\\ }
\maketitle

\begin{abstract}
  We present new \xmm\ data of the high-redshift ($z=1.883$), \mpc
  -sized giant radio galaxy \qc. The larger collecting area and longer
  observation time for our new data means that we can better
  characterise the extended X-ray emission, in particular its spectrum,
  which arises from cosmic microwave background photons scattered into
  the X-ray band by the energetic electrons in the spent synchrotron
  plasma of the (largely) radio-quiet lobes of \qc. We calculate the
  energy that its jet-ejected plasma has dumped into its surroundings
  in the last $3 \times 10^7$ years and discuss the impact that
  similar, or even more extreme, examples of spent, radio-quiet lobes
  would have on their surroundings. Interestingly, there is an
  indication that the emission from the hotspots is softer than the
  rest of the extended emission and the core, implying it is due to
  synchrotron emission. We confirm our previous detection of the
  low-energy turnover in the eastern hotspot of \qc.
\end{abstract}


\begin{figure*}
\rotatebox{0}{
\resizebox{!}{8cm}
{\includegraphics{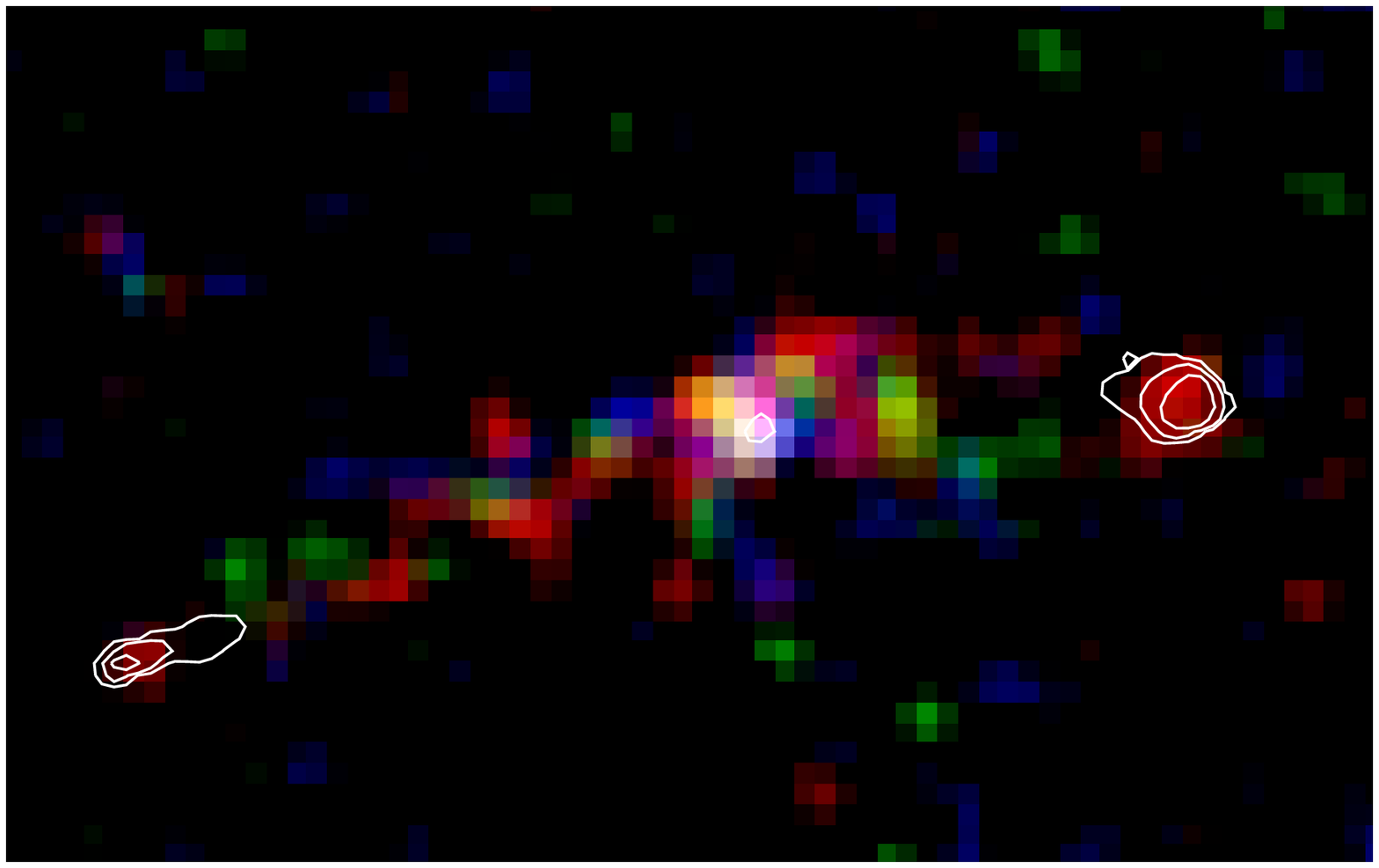}}}
\caption{Image of our \xmm\ data ($132''\times 83''$ with $2$\as\
  pixels; the image has been smoothed by a Gaussian with $\sigma$ of
  $2$ pixels, the core of \qc\ is located at RA 09h08m16.9s, Dec
  +39d43m26s in J2000). The three colours represent soft ($0.2 -
  1$\kev\ band; red), medium ($1-3$\kev\ band; green) and hard
  ($3-10$\kev\ band; blue) X-ray bands. Overlaid in white are VLA
  A-array $1.425$-\ghz\ radio contours ($0.4, 1.7$ and $7$\mjypb). The
  X-ray emission from the hotspots is softer than the lobes and core
  and so they appear red.}
\label{fig:colour}
\end{figure*}

\section{Introduction}

At redshift $z=1.88$, \qc\ is one of the highest redshift giant radio
galaxies known.  It spans $111$\as\ on the sky \citep{lawgreen95},
which represents a projected size of $945$\kpc\ in the cosmology
assumed in this paper.  \citet{6c0905} observed this powerful,
classical double \frii\ \citep{FR} radio galaxy with \chandra\ and
detected extended X-ray emission along the axis of the radio source
which we concluded was emitted by inverse-Compton scattering of the
Cosmic Microwave Background (ICCMB, CMB) by the lobes (undetected in
the radio) of the radio galaxy.  The sheer physical size of the source
means that the extended emission cannot be due to up-scattering of a
nuclear photon component and the breadth of the X-ray lobes rules out
jet emission via beamed ICCMB. It is common for the lobes of radio
sources at low- and high-redshift to emit X-rays in this way
(e.g. \citealt{croston05}, \citealt{overzier05}, \citealt{erlund06}).
Higher redshift objects are easier to detect and/or less ambiguous
than low-redshift objects because ICCMB is not redshift-dimmed
compared with other emission processes \citep{schwartz02}. The energy
stored in ICCMB-scattering electrons can potentially be considerable
and is more representative of the actual amount of energy that a radio
source pumps into its environment than synchrotron radio emission
because it is much longer lived and probes the lower Lorentz-factor
electrons thought to dominate typical particle energy distributions.

In this paper we present recent \xmm\ observations of \qc.
Throughout this paper, all errors are quoted at $1\sigma$ unless
otherwise stated and the assumed cosmology is $\rm H_{\rm 0} =
71$\kmpspmpc, $\Omega_{0}=1$ and $\Omega_{\Lambda} = 0.73$.  One
arcsecond represents $8.518$\kpc\ on the plane of the sky at the
redshift ($z=1.883\pm 0.003$) of \qc\ and the Galactic absorption
along the line-of-sight is $1.91 \times 10^{20}$\pcmsq\
\citep{dickeylockman90}.


\section{Data Reduction}
\label{sec:reduction}

Our \xmm\ data of \qc\ consists of $58.0$\ks\ of EPIC-pn data and
$61.9$\ks\ of data for each of the EPIC-MOS cameras. \qc\ was observed
on 2006 October 30. The standard pipeline was used to reduce the data,
SAS tools EPCHAIN and EMCHAIN were used for the EPIC-pn and both
EPIC-MOS data respectively. After filtering the resulting files to
remove periods dominated by flares and taking dead-time intervals into
account, $42.2$\ks\ of good-time for the PN, and $53.1$\ks\ for both
MOS\,1 and MOS\,2 were left. In total, in the $0.2-8$\kev\ band (after
summing the MOS\,1, MOS\,2 and the PN data and using the regions shown
in Fig \ref{fig:regs}), there are 1271 counts (of which 579 are
background) in the source, 830 (422 background) in the extended
emission (lobes and hotspots) and 441 counts (159 background) in the
core. Spectra for the extended X-ray emission (excluding the 15\as\
core region), the nucleus ($15$\as\ circle) and background (an area of
sky free from sources near \qc\ on the same chip) were extracted
separately for each instrument. They were then stacked and fitted
using \xspecv. All spectral data were fitted over the $0.5-10$\kev\
band. The nucleus of \qc\ is addressed in \citet{core} and so is not
discussed further here.

The radio data used in this paper first appeared in \citet{6c0905} and
were reduced using standard AIPS techniques.

\section{Results}

Our \xmm\ data clearly show X-ray emission extending along the radio
axis of \qc\ to the east and west (Fig. \ref{fig:colour}). This was
first detected by \chandra\ and presented in \citet{6c0905} where we
noted that, as \chandra\ resolves the transverse extent of this
extended emission ($\sim 40$\kpc, although the paucity of counts makes
the difficult to measure this accurately), it cannot be jet emission
(since this would be very narrow) but rather is due to inverse-Compton
up-scattering of the CMB by spent radio plasma from the lobes. The
\xmm\ data supports this interpretation. The spectra extracted from
the EPIC-pn, MOS\,1 and MOS\,2 data were binned by 20 counts per bin
and associated with a background spectrum, response and ancillary
response files using GRPPHA.  They were then fitted with a
Galactic-absorbed power-law using XSPEC (which subtracts the
background) giving a photon index of $\Gamma=1.61^{+0.19}_{-0.17}$ and
a normalisation of $1.9^{+0.2}_{-0.2}\times 10^{-6}$\photpkevpcmsqps\
with a reduced-chi-squared of $\chi^2_\nu = 0.98$ with $31$ degrees of
freedom (d.o.f.). The absorption-corrected X-ray luminosity of the
extended emission in the $2-10$\kev\ band is $L_{\rm X} =
1.5^{+0.1}_{-0.2}\times 10^{44}$\ergps, and the observed flux in the
same band is $F_{\rm X} = 9^{+2}_{-2}\times 10^{-15}$\ergps.
\citet{6c0905} found $\Gamma = 2.7^{+1.7}_{-1.7}$ from the \chandra\
data, but the \xmm\ spectra are better constrained and typical of
extended inverse-Compton emission associated with radio galaxies
(e.g. \citealt{croston05}; \citealt{erlund06}).  (Note that the values
presented for the extended emission in \citealt{core} came from a
larger extraction region.)

\begin{figure}
\rotatebox{0}{
\resizebox{!}{8cm}
{\includegraphics{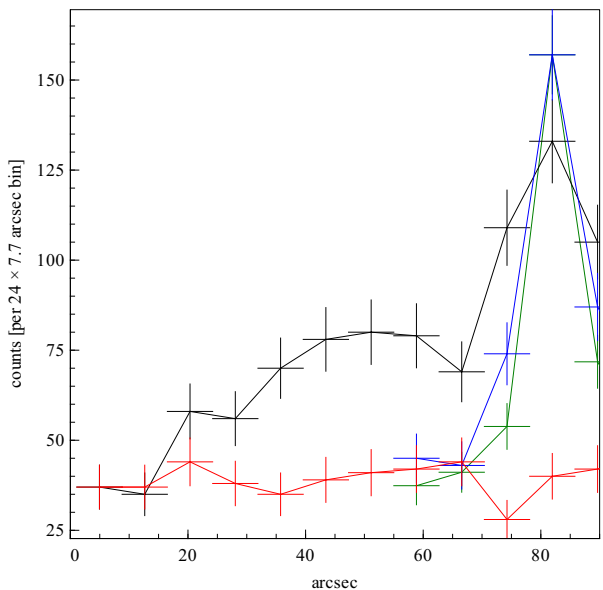}}
}
\caption{Profiles taken from the $0.2-8$\kev\ mosaic of PN, MOS1 and
  MOS2 images.  Each point represents a $7.7 \times 24$\as\ bin. The
  black line represents a profile along the eastern lobe of \qc\
  only. The blue line is a profile going from north-east to south-west
  through the core.  The red line represents a profile of a
  source-free region on the same chip. The green line represents a
  slice through a bright source to the north-east of \qc, which has
  been normalised so that it matches the peak of the core profile. }
\label{fig:profile}
\end{figure}

The number of X-ray photons increases towards the core (by $\sim 50$
per cent) as can be seen in Fig. \ref{fig:colour} and Fig.
\ref{fig:profile}. The latter figure shows four profiles: the eastern
lobes in black; a source-free region in red to give an idea of the
background level on the chip; the core in blue (i.e. roughly
perpendicular to the radio axis); and a bright (comparison) source to
the north-east of \qc\ which has been normalised so as the sides of
the core profile are roughly matched in green. This green line
illustrates that the point spread function (PSF) of the central core
does not significantly contaminate the extended emission.  Each
profile is calculated using a rectangular region 24\as\ wide from a
stacked $0.2-8$\kev\ PN, MOS1 and MOS2 image with $7.7$\as\ pixels.
The PSF is $\sim 7.4$ arcsec at the FWHM which corresponds to one
point.

\begin{figure*}
\rotatebox{0}{
\resizebox{!}{8cm}
{\includegraphics{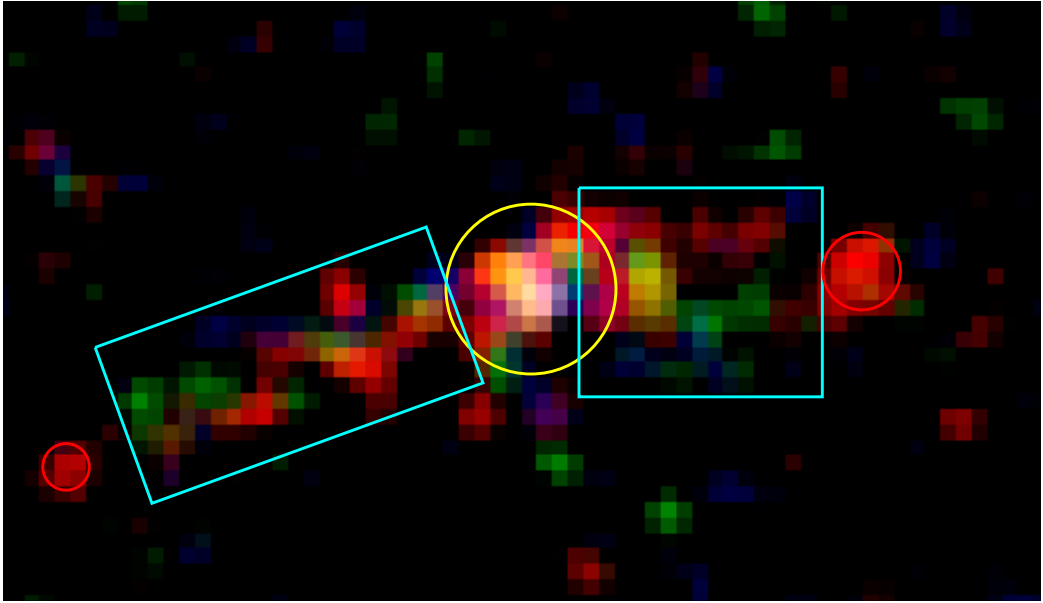}}
}
\caption{Three colour image ($132''\times 77''$; energy bands, pixels
  sizes and smoothing as in Fig. \ref{fig:colour}) illustrating the
  regions that were used when calculating the hardness ratio in the
  hotspots, lobes and core. The core region was excluded from the lobe
  regions.  In Fig.  \ref{fig:hr}, region 2 is the core region
  depicted here by a yellow circle, regions 1 and 3 are the lobe
  regions indicated as cyan rectangles and the hotspot regions, shown
  by small red circles, are regions 0 and 4.  It can clearly be seen
  that the hotspot regions mainly contain soft (red) photons, in
  contrast with the other regions.  }
\label{fig:regs}
\end{figure*}

\begin{figure}
\rotatebox{0}{
\resizebox{!}{8cm}
{\includegraphics{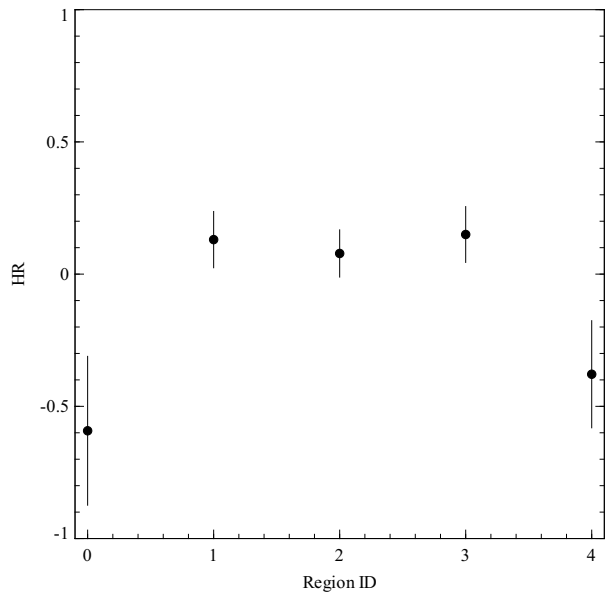}}
}
\caption{Hardness ratios calculated for the core (region 2); the two
  lobe regions (regions 1 and 3) and the two hotspot regions (regions
  0 and 4), as illustrated in Fig.  \ref{fig:regs}.  The hardness
  ratio was calculated using the soft ($0.2-1$\kev) and hard
  ($1-8$\kev) bands.  The error bars represent one standard deviation
  and are calculated using a Monte-Carlo simulation to propagate the
  errors on the two energy bands and due to the background.  This plot
  shows that the hotspots are softer than the lobes and nucleus.}
\label{fig:hr}
\end{figure}

\begin{figure}
\rotatebox{0}{
\resizebox{!}{8cm}
{\includegraphics{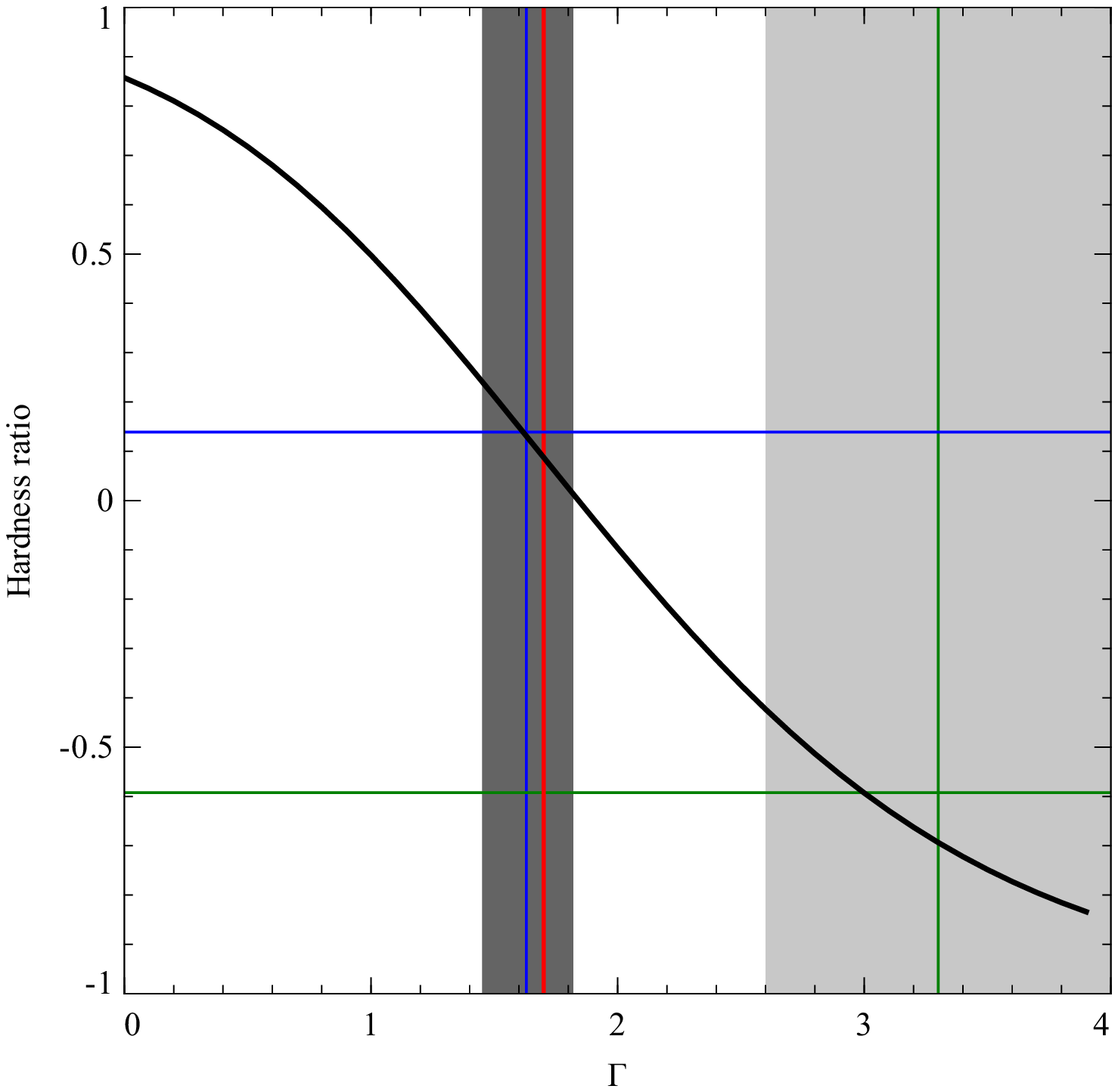}}
}
\caption{The thick solid black curve shows how hardness ratio (HR)
  varies as a function of photon index ($\Gamma$). The green lines
  represent the measured hardness ratio and photon index (from
  spectral fitting) of the eastern hotspot region. The blue lines
  represent the same quantities for the lobe regions.  The dark and
  light grey boxes represents the error on the photon index of the
  lobe and the eastern hotspot respectively from spectral fitting. The
  plot shows that the hardness ratio of the eastern hotspot in the
  X-ray is lower, so softer, than the radio hotspots. The thin red
  line represents the radio photon index considering the integrated
  flux density of the eastern radio hotspot and small radio emitting
  lobes together (calculated from values given in \citealt{lawgreen95}
  and shown outlined by white contours in Fig. \ref{fig:colour}).}
\label{fig:GammaHR}
\end{figure}

\citet{6c0905} noted that most of the X-ray emission in the \chandra\
observation lay between the radio hotspots and core, along the lobes
of \qc. The more extended western hotspot emitted in X-rays but the
eastern one did not at the upper limit of our \chandra\
observation. In \citet{6c0905}, we interpreted the lack of X-ray
emission from the eastern hotspot together with the presence of X-ray
emission in the lobe material nearby as a direct detection of the
low-energy turnover in the electron population ($\gamma_{\rm
  min}$). Our deep \xmm\ data show a very similar picture, with rather
different X-ray emission associated with the eastern hotspot from that
of the rest of the associated lobe. Interestingly, our new \xmm\
observation reveals some very soft photons (approximately two thirds
of these photons have energies below $0.8$\kev) associated with the
eastern hotspot, which would expect to be undetectable in our current
\chandra\ observation.  We consider the nature of this very soft
emission in Section \ref{sec:gmin}.

In order to determine whether there is any significant difference
between the hardness ratio\footnote{Here we calculate the hardness
  ratio, $HR = \frac{H-S}{H+S}$, where $H$ is the number of counts in
  the hard band and $S$ is the number of counts in the soft band} of
the X-ray emission from the hotspots, lobes and core, the number of
counts in the regions shown in Fig.  \ref{fig:regs} were calculated
for the PN and both MOS cameras in the soft and hard band
($0.2-1$\kev\ and $1-8$\kev) respectively using {\sc CIAO} tool {\it
  dmstat}.  The counts in each region were calculated separately and
by careful specification of the region files within {\it ds9}
contamination of the core region in the lobe region was avoided.
These energy bands were chosen so that the number of counts in each
band would be roughly equal.  The hardness ratio in each region was
calculated taking into account the background counts in each band.
The resulting hardness ratio profiles as illustrated in
Fig. \ref{fig:hr}, which shows that both hotspots are softer than the
lobes and nucleus.  A constant hardness ratio model is a poor fit to
the data giving a best fit hardness ratio of $0.08$ and a $\chi^2_\nu
= 2.6$.  Fig.  \ref{fig:profile} shows that the difference between the
hardness ratio of the lobes and hotspots is unlikely to be due to
contamination from the core.

The relationship between the X-ray photon index and hardness ratio is
plotted in Fig. \ref{fig:GammaHR}. \xspec\ was used to evaluate this,
taking into account the Galactic absorption along the line-of-sight to
\qc. Fig. \ref{fig:GammaHR} shows that lower hardness ratios imply
steeper spectra. Thus at X-ray energies, the lobes have a flatter
spectrum than the hotspots. 

\section{discussion}
\label{sec:discussion}


The diffuse X-ray emission spanning the gap between the core and the
radio-emitting hotspots in \qc\ is a clear case of inverse-Compton
scattering of CMB photons by the relativistic electrons stored in the
spent synchrotron plasma once accelerated in the hotspot (as per the
standard model of \frii\ radio galaxy evolution see
e.g. \citealt{blundell99}).  Beamed ICCMB emission from a relativistic
jet can be ruled out from the morphology of the X-ray emission in the
\chandra\ data which, at $\sim 40$\kpc, is too laterally extended to
correspond to a jet \citep{6c0905}.

\subsubsection{The X-ray emission mechanism in the hotspots}

The difference in the X-ray hardness ratio (and correspondingly the
spectral index / photon index) between the hotspots and the lobes
emission implies that different X-ray emission processes are taking
place.  \citet{lawgreen95} studied the spectral indexes of the radio
emission associated with the hotspots and the small amount of lobe
emission detected.  The radio spectral indexes\footnote{The spectral
  index, $\alpha$, is related to the photon index, $\Gamma$ as $\alpha
  = \Gamma - 1$, $\alpha$ is defined as $S_\nu \propto
  \nu^{-\alpha}$.} of the eastern and western hotspots are $\alpha =
0.8$ and $1.0$ respectively, whereas the small radio lobes associated
with these hotspots have spectral indexes of $\alpha = 1.3$ and $1.7$
respectively (these values come from \citealt{lawgreen95} who do not
provide errors). In other words in the radio the spectral index may
steepen from the hotspot to the lobe as would be expected from a
curved particle distribution in a region with a magnetic field
gradient (for reasons why this model is most likely see
\citealt{blundellrawlings00}). If it were the case that the X-rays
from both the hotspot and lobes were from ICCMB emission, then the
same curved spectrum and magnetic field should also cause the X-ray
ICCMB emission to steepen between the hotspot and lobes (because the
energy distribution will shift on expansion from the hotspot to the
lobes). In fact the X-ray emission is steeper (i.e. softer) in the
hotspot than in the lobes, this and the steepness of the spectrum
implies that we are seeing the high energy tail of synchrotron
emission in the hotspot.

Synchrotron self-Compton (SSC) emission has the same spectral shape as
the synchrotron spectrum that has been up-scattered. The spectral
indices of the radio hotspots are flatter than the X-ray spectral
indices which would argue against this emission mechanism in the
hotspot. However, to calculate the expected SSC flux in the eastern
hotspot, we have assumed a minimum energy magnetic field with no
low-energy cut-off ($B_{\rm min} = 2.2\times 10^{-5}$\g, for
$\gamma_{\rm min} = 1$), that the 408\mhz\ emission ($107$\mjy, peak
and lobe flux; \citealt{lawgreen95}) provides the SSC seed photon
field and that the region responsible for up-scattering photons into
the X-ray band is cylindrical with a diameter of 28\kpc\ and a length
of 56\kpc.  The predicted SSC 1\kev\ X-ray flux density is then
$7\times 10^{-5}$\njy.  This is several orders of magnitude lower than
what we actually detect, which is $\sim 0.3$\njy\ (calculated from a
power-law fitted to the eastern and western hotspot where the
normalisation between the two regions was independently
fitted). However, \citet{6c0905} inferred a low-energy turnover of
$\gamma_{\rm min} \gtapprox 10^4$ in the population of this hotspot, in
which case $B_{\rm min} = 8.0\times 10^{-6}$\g\ and the predicted
1\kev\ flux is $3.8\times 10^{-4}$\njy. However, this would mean that
the electrons emitting radio synchrotron lie below the low-energy
turnover and so this emission mechanism can be ruled out (for
confirmation of the detection of the low energy turnover in this
object see Section \ref{sec:gmin}).

\begin{figure}
\rotatebox{0}{
\resizebox{!}{8cm}
{\includegraphics{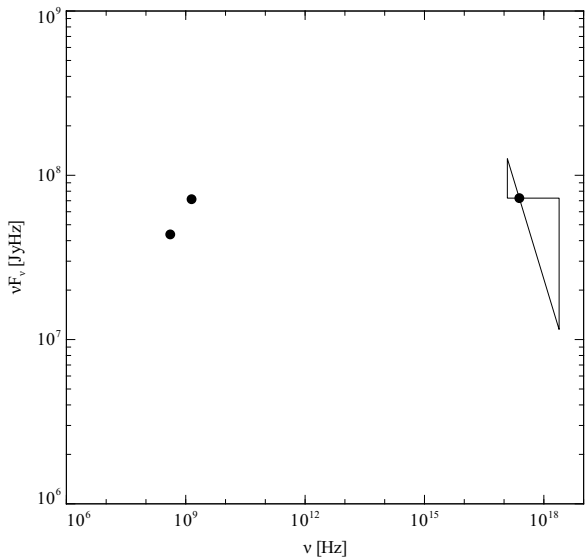}}
}
\caption{The SED of the eastern hotspot of \qc\ showing radio and
  X-ray data points.  Radio points are integrated values for the
  eastern radio component (peak and lobes) and are taken from
  \citet{lawgreen95}.  The spectral index calculated from these values
  between 408\mhz\ and 1.4\ghz\ is $\alpha \sim 0.7$ which is slightly
  flatter than $\alpha \sim 0.8$ quoted in \citet{lawgreen95}.  The
  X-ray point is calculated from the normalisation of the eastern
  hotspot when the eastern and western hotspot are fitted together
  giving a normalisation of $4.2^{+2.0}_{-1.6}\times
  10^{-7}$\photpkevpcmsqps.  This gives a photon index of $\Gamma =
  2.4\pm 0.4$ averaged over the eastern and western hotspot regions.
  This is a conservative measure of the steepness of the eastern
  hotspot as the harder component present in the western hotspot
  flattens the combined spectrum.  However, it is better constrained
  than either the spectral fit of the eastern hotspot on its own or
  the photon index that we can infer from the hardness ratio of the
  eastern hotspot (see Fig. \ref{fig:hr} and Fig. \ref{fig:GammaHR})}
\label{fig:sed}
\end{figure}

Thus we conclude X-ray synchrotron emission from the hotspots, where
the magnetic field strength is likely to be enhanced compared with
that of the lobes and where particles are being accelerated, is the
most likely explanation for the change in spectral index between the
hotspot and the lobe emission. The 1\kev\ X-ray flux density is
consistent with a cooling or curved synchrotron spectrum (see Fig.
\ref{fig:sed}). Extrapolating the radio flux density at 408\mhz\
(107\mjy) up to 1\kev\ (assuming a straight power-law particle
distribution) using the radio spectral index for the eastern hotspot
($\alpha \sim 0.7$, calculated from the integrated flux density values
for the eastern hotspot--lobe component given in \citealt{lawgreen95})
gives a predicted monochromatic X-ray flux density of $\sim 80$\njy;
we detect $\sim 0.3$\njy, consistent with a steepening particle
distribution curved in the usual way.

\subsubsection{The electron population in the hotspots}
\label{sec:gmin}

Noting the eastern hotspot of \qc\ to be inefficient at up-scattering
CMB photons into the X-ray band, yet the nearby lobe to be efficient,
\citet{6c0905} deduced that this was direct evidence of a low-energy
turnover in the electron population. The morphology of our new \xmm\
data is similar to that of our \chandra\ data.

Assuming a power-law electron distribution which extends back to low
Lorentz factors ($\gamma_{\rm min} \sim 1$), we make use of Equation 1
in \citet{celottifabian04} to determine the ratio of the predicted
monochromatic X-ray luminosity, $L_{\rm X}$, to monochromatic radio
luminosity, $L_{\rm R}$ as follows
\begin{equation}
\frac{L_{\rm X}}{L_{\rm R}} = \frac{\mathcal{U}_{\rm CMB}}{\mathcal{U}_{\rm B}}
\left(\frac{\nu_{\rm X}\nu_{\rm B}}{\nu_{\rm R}\nu_{\rm CMB}}\right)^{1-\alpha}
(1+z)^{(3+\alpha)}
\end{equation}
for the eastern hotspot, where $\mathcal{U}_{\rm CMB}$ is the energy
density of the CMB at redshift $z=0$ and $\mathcal{U}_{\rm B}$ is the
energy density of the magnetic field. $\nu_{\rm X}$, $\nu_{\rm R}$ and
$\nu_{\rm CMB}$ are the observed frequencies in the radio and X-ray
and the frequency of the peak of the CMB (again, at $z=0$) and
$\nu_{\rm B}$ is the gyro-frequency.  $\alpha$ is the spectral index
of the radio emission.  Re-writing Equation 1 to express it as a
function of magnetic field strength, $B$, gives
\begin{equation}
\frac{L_{\rm X}}{L_{\rm R}} = 2^{2+\alpha}\pi^{\alpha}(1+z)^{3+\alpha}\mathcal{U}_{\rm CMB}
\left(\frac{e}{m_{\rm e}c}\frac{\nu_{\rm X}}{\nu_{\rm R}\nu_{\rm CMB}}\right)^{1-\alpha}
B^{-(1+\alpha)}
\end{equation}
where $e$ is the charge on the electron, $m_{\rm e}$ is the mass of
the electron and $c$ is the speed of light.

\begin{figure}
\rotatebox{0}{
\resizebox{!}{8cm}
{\includegraphics{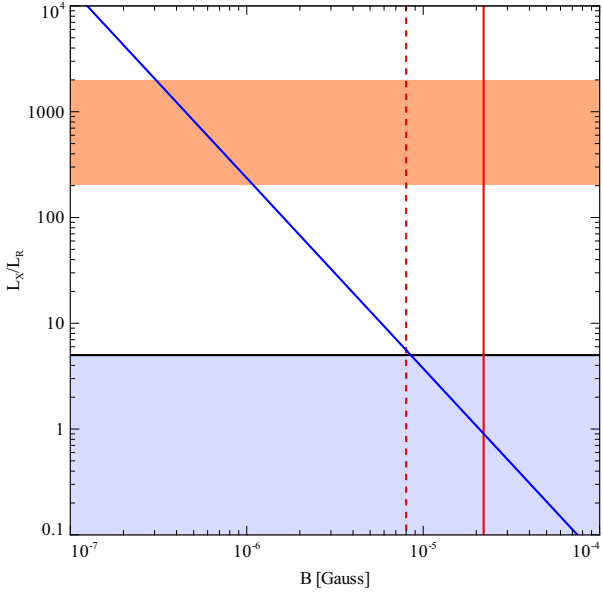}}
}
\caption{The plot shows the dependence of $\frac{L_{\rm X}}{L_{\rm
      R}}$ on the magnetic field. If the very soft X-rays we detect in
  the eastern hotspot are due to synchrotron emission then this would
  provide a lower limit on the detectability of the $1$\kev\ X-ray
  luminosity due to ICCMB (solid black line representing the observed
  1\kev --408\mhz\ monochromatic luminosity ratio). Below this ratio,
  in the blue shaded region, the X-ray luminosity predicted would be
  undetectable.  The solid blue line represents the predicted
  $\frac{L_{\rm X}}{L_{\rm R}}$, using the 408\mhz\ radio luminosity,
  as a function of $B$-field. The solid red line is the minimum energy
  magnetic field ($B_{\rm min}$) in the eastern hotspot calculated
  assuming a cylindrical emission region with a diameter of 28\kpc\
  and a length of 56\kpc, using the radio flux detected at 408\mhz\
  and assuming $\gamma_{\rm min} =1$. This intersects with the
  predicted $\frac{L_{\rm X}}{L_{\rm R}}$ line in the undetectable
  region. Similarly, the dashed red line shows $B_{\rm min}$ assuming
  $\gamma_{\rm min} = 10^4$.  The orange region illustrates the range
  in $\frac{L_{\rm X}}{L_{\rm R}}$ deduced from the X-ray bright lobe
  emission.  The range in the orange region comes from the uncertainty
  in the length of the hotspot (here we estimate it to be between a
  tenth to a hundredth of the length of the lobe).  This region lies
  well above the undetectable (blue shaded) region clearly showing
  that there must be a low-energy turnover, because the $\gamma \sim 10^3$
  electrons we would expect to find in the eastern hotspot are not
  there.}
\label{fig:magdep}
\end{figure}

Fig. \ref{fig:magdep} shows the dependence of the X-ray--radio
luminosity ratio on magnetic field strength at the redshift of \qc: as
the magnetic field decreases, $\frac{L_{\rm X}}{L_{\rm R}}$ increases.
In order to detect ICCMB emission from the eastern hotspot, the
magnetic field must be well below the minimum-energy magnetic field,
$B_{\rm min}$, assuming no low-energy turnover.  We note, however,
that $B_{\rm min}$ depends on $\gamma_{\rm min}$, with a higher
$\gamma_{\rm min}$ resulting in a lower $B_{\rm min}$ (as described in
\citealt{6c0905} and illustrated in Fig. \ref{fig:magdep}). It is
therefore not possible to determine whether there is a low-energy
turnover in the eastern hotspot using just our \xmm\ and radio hotspot
data so we now consider the other relevant information from these
observations.

In \citet{6c0905} we considered the implications of the X-ray bright
lobes and X-ray faint hotspots in \qc.  We concluded that the X-ray
emission from the lobes implied that there must be a low-energy
turnover in the eastern hotspot.  The same argument still
holds. Assuming a constant hotspot advancement speed (and that any
expansion due to cooling takes place laterally), we can consider a
region of lobe emission of the same length as the current eastern
radio hotspot. The electrons responsible for inverse-Compton
scattering CMB photons in the lobe have a Lorentz factor of $\gamma
\sim 10^3$. These electrons will have cooled by at least an order of
magnitude \citep{blundell99} since they left the hotspot.  So
calculating the number of $\gamma \sim 10^3$ electrons in a
hotspot-sized region of the lobe tells us roughly the number of
$\gamma \sim 10^4$ electrons there must have been in the hotspot.  We
can therefore calculate the number of $\gamma \sim 10^3$ electrons
that there would be in the hotspot if there were no low-energy
turnover, for an assumed spectral index.  Then we calculate how much
ICCMB emission these $\gamma \sim 10^3$ electrons would have produced.
The ratio between this and the radio emission at 408\mhz\ is shown as
an orange region in Fig. \ref{fig:magdep} (the range of $\frac{L_{\rm
    X}}{L_{\rm R}}$ reflects the uncertainty in the size of the
hotspot region.  We consider the lobe-length--to--hotspot-length ratio
to be between 10 and 100.)  Fig. \ref{fig:magdep} clearly shows that
there must be a high low-energy turnover in the hotspot of \qc\ with a
$\gamma_{\rm min} \gtapprox 10^4$ as the expected $1$\kev\ X-ray luminosity
is two orders of magnitude above our detection limit.

Fig \ref{fig:magdep} implies that if we are seeing the low-energy
turnover then the magnetic field in this hotspot must be relatively
weak.  This could explain why we see X-ray synchrotron hotspot
emission from this source when usually high-luminosity \frii\ radio
galaxies do not have high frequency synchrotron emission associated
with their hotspots.  This is thought to be because the magnetic field
is too strong (see, for example, \citealt{meisenheimer89} and
\citealt{blundell99}).

We assume that power is transferred from the central black hole to the
hotspots of 6C0905 along jets with high efficiency and with little
radiation, presumably because most of the transferred energy is
kinetic in the form of bulk motion of particles.  Although a high bulk
Lorentz factor $\Gamma_{\rm b}\sim 10$ could make radio through X-ray
radiation from $\gamma\sim 10^{3-5}$ particles in the jets
undetectable by relativistic beaming, it would also greatly amplify
the energy density of the CMB in the jet frame.  This would lead to
steep inverse-Compton losses of such energetic electrons, so electrons
with these energies come from re-acceleration at the hotspots.

\subsubsection{Increase in X-ray lobe photons towards the core}
\label{sec:increase}

There is a $\sim 50$ per cent increase of X-ray emission towards the
core (see Fig. \ref{fig:profile}; 30\% of the total extended emission
contaminates the core spectrum; see \citealt{core} for details).
Given that the X-ray emission process is inverse-Compton, this implies
that there are either more electrons with $\gamma \sim 10^3$ or that
the energy density of the seed photon field has increased.

The central galaxy is bright in the infra-red ($\nu L_\nu \sim 4\times
10^{44}$\ergps, at 5.8\um; \citealt{seymour07}\footnote{calculated
  using the flux density in the IRAC 5.8\um\ band}) so will provide a
source of photons which could be up-scattered (off lower Lorentz
factor electrons, $\gamma \le 100$) along with the CMB (off $\gamma
\sim 10^3$ electrons); however, this photon field will be dominated by
the CMB beyond $\sim 0.5$\as\ of the central galaxy\footnote{The
  energy density of the photons is calculated using $\mathcal{U}_{\rm
    IR} = \frac{L_{\rm IR}}{\pi c R^2}$, equating it to the energy
  density of the CMB which is $\sim 3 \times 10^{-11}$\ergpcmc\ and
  solving for $R$, which is the distance from the nucleus in cm.  The
  galaxy has been approximated as a point source.} and so can be ruled
out.

The amount of ICCMB emission in the X-ray depends only on the number
density of the electrons with $\gamma\sim 10^3$ as the energy density
of the CMB is constant throughout the source (and is independent of
the magnetic field strength). One way in which there could be
increasing numbers of $\gamma\sim 10^3$ electrons closer to the core
is if we are seeing the low-energy turnover in the electron population
as the back-flowing plasma approaches the core and cools. There could
be a contribution from more lobe plasma closer to the nucleus.

\subsubsection{Feedback}

Emission that arises from inverse-Compton scattering of the CMB probes
electrons with a lower energy than those typically responsible for
producing radio synchrotron emission. In the case of \qc, it means
that the lobes of this high-redshift giant radio galaxy are detected
in X-rays but not in the radio. Using the long-lived X-ray emission we
can probe the amount of energy that \qc\ has pumped into its
environment in the last $\sim 3\times 10^7$\yr, $\mathcal{E}_{\rm e}$:
approximately the length of time required to cool the electrons
responsible for up-scattering the CMB into the X-ray band i.e those
with Lorentz factors $\gamma \sim 1000$.  We follow the arguments in
\citet{erlund06} making use of their Equation 4: \begin{equation}
  \mathcal{E}_{\rm e} = \frac{3}{4}\frac{L_{\rm X} m_{\rm e} c}
  {\mathcal{U}_{\rm rad}\gamma_{\rm e}\sigma_{\rm T}} \simeq
  \frac{3}{4}\frac{L_{44}}{\gamma_{\rm e}(1+z)^4}10^{64},
\end{equation}
where $L_{\rm X}$ is the X-ray luminosity in the $2-10$\kev\ band and
$L_{\rm X} = L_{44} \times 10^{44}$\ergps. The luminosity of the
extended emission (both lobes and hotspots) emission is $L_{\rm X} =
1.5^{+0.1}_{-0.2}\times 10^{44}$\ergps\ in the $2-10$\kev\ band, but
there is a further $L_{\rm X} \sim 1.0\times 10^{44}$\ergps\
contaminating the core spectrum \citep{core}, meaning that the whole
of the extended emission has a luminosity of $L_{\rm X} \sim 3\times
10^{44}$\ergps. $\mathcal{U}_{\rm rad}$ is the energy density of the
photon field which is being up-scattered, in this case the
CMB. $\gamma_{\rm e}$ is the Lorentz factor of the electrons
responsible for most of the up-scattering: we assume that $\gamma_{\rm
  e} = 1000$.  $\sigma_{\rm T}$ is the Thomson cross-section. The
lower limit to the amount of energy pumped into the environment by
\qc\ is $\sim 3\times 10^{59}$\erg; including protons can increase
this by two to three orders of magnitude to $\sim 10^{62}$\erg\ or
more.

Some clusters of galaxies require a large amount of energy deposited
into the X-ray emitting gas to offset cooling and explain why
significant amounts of cool gas are not detected in the X-ray in the
centres of these systems. This feedback requires a power input of the
order of $L_{\rm feedback}\sim 10^{44\pm1}$\ergps\ \citep{dunn06}
which gives a total energy of $\mathcal{E}_{\rm feedback}\sim
10^{59\pm 1}$\erg\ over $3\times 10^{7}$\yr.  This simple comparison
is to illustrate the huge amount of energy stored in the lobes of
powerful radio galaxies: enough to stem a cooling flow. Other powerful
\frii\ radio sources have similar lower limits to their reservoirs of
energy stored in electrons (i.e. \citealt{erlund06}).  Most radio
galaxies in clusters are \fri\ radio sources which are likely to be
less powerful.

\subsubsection{Energy in ghost reservoirs}

\citet{saripalli05} found four giant radio galaxies at $z<0.13$ in a
statistically complete survey using the 2100\,deg$^2$ SUMSS field.
None of these sources were classical double \frii\ radio galaxies.
One is an \fri\ and the other three are also below the \fri --\frii\
luminosity divide but are possibly relic \frii\ sources because they
have relaxed lobes and no hotspots (see also \citealt{cordey87} for
another example of a large relic radio galaxy). This illustrates how
rare sources such as \qc\ are at low redshifts.  If three of
\citet{saripalli05} relaxed doubles are in fact relic sources as they
suggest, then powerful galaxies will be more common at higher redshift
(closer to the peak in the space density of radio sources). \qc\ may
be such an example as it has clear hotspots and, unlike its lower
redshift counterparts, it has very little lobe emission at radio
frequencies \citep{lawgreen95}.  The lack of radio lobe emission is
due to a lack of particles having the right energies to radiate
synchrotron emission at radio wavelengths, given the ambient magnetic
field strength. (Radio observations at 408\mhz, \citealt{lawgreen95},
and at 1.4\ghz, \citealt{6c0905}, appear not to have suffered from
flux loss due to interferometric under-sampling, see
\citealt{6c0905}.)

Sources such as the three relaxed doubles in the \citet{saripalli05}
complete sample contain large reservoirs of aged synchrotron plasma
which is ideally placed to up-scatter the CMB.  These sources appear
to be the most common sort of giant radio galaxy at low redshift based
on the results in \citet{saripalli05} (they made up three out of four
of the sources in their complete sample).  Most giant
radio galaxies could contribute to the anisotropies in the CMB and
therefore are a potential contaminant in SZ (Sunyaev-Zel'dovich)
surveys, particularly relic giant radio sources which are not strong
radio emitters and so would be undetectable in the radio at relatively
low redshifts.  Given that most of the X-rays associated with \qc\ are
associated with its extended emission, and are not co-spatial with the
radio emission, this radio galaxy has pumped an otherwise completely
undetectable $\sim 3\times 10^{59}$\erg\ into its environment.


\section{conclusions}
\label{sec:conclude}

Our \xmm\ observation of the giant, high-redshift ($z=1.88$) radio
galaxy \qc\ clearly shows extended X-ray emission aligned along the
radio axis of the source. The longer observation time and greater
sensitivity of our \xmm\ observations means that we can better
characterise this emission. It is richer than the previous \chandra\
data, with an increase in X-ray emission towards the central source
and a difference in the hardness ratio of the hotspots and lobe
emission. We demonstrate that the soft X-ray emission in the hotspots
of \qc\ is most likely to be X-ray synchrotron emission. We confirm
our previous detection of the low-energy turnover in the eastern
hotspot and suggest that the increase in X-ray emission towards the
core is due to the low-energy turnover in the electron population
being revealed as the electron population cools.

We find that the minimum energy that \qc\ has pumped into its
environment in the last $\sim 3\times 10^7$\yr\ is $\sim 3 \times
10^{59}$\erg\ (protons could increase this by a factor of $10^2 -
10^3$) and that it is clear that in \qc, older plasma is more
efficient at up-scattering the CMB than freshly accelerated plasma.
This implies that there are more electrons with $\gamma \sim 10^3$ in
the older plasma and thus that the younger plasma has a low-energy
turnover above $\gamma_{\rm min} \sim 10^3$.

The complete sample of \citet{saripalli05} of low-redshift giant radio
galaxies shows that actively fed giant \frii\ sources are rare (they
found none), but that relic \frii\ sources may be the most common type
of low-redshift giant radio galaxy (three out of four of their sources
were potential relic sources). The Universe may be littered with the
X-ray bright relics which could in their turn have an effect on the
detected CMB anisotropies and hence SZ surveys. It is important to
note that these would not be correlated with the observed distribution
of currently radio-bright extended sources.

\section*{Acknowledgements}

MCE acknowledges STFC for financial support. ACF and KMB thank the
Royal Society. 

\bibliographystyle{mnras} 
\bibliography{mn-jour,6c0905_xmm_lobes}
\end{document}